\documentclass[review]{elsarticle}
\usepackage[utf8x]{inputenc}
\usepackage{graphicx}
\usepackage{amsmath}
\usepackage{amssymb}
\usepackage{lineno,hyperref}
\modulolinenumbers[5]
\journal{Journal of Systems and Software}
\usepackage{color}
\usepackage{multirow}
\newcounter{qcounter}
\newcounter{acounter}

\begin{document}

\begin{frontmatter}

\title{Extracting State Transition Models from i$^*$ Models}

\author[univ1]{Novarun Deb\corref{corrauth}}
\cortext[corrauth]{Corresponding author}
\ead{novarun.db@gmail.com, M:+91-9830715623}

\author[univ1]{Nabendu Chaki}
\ead{nabendu@ieee.org}

\author[univ2]{Aditya Ghose}
\ead{aditya@uow.edu.au }

\address[univ1]{Department of Computer Science and Engineering, University of Calcuttta, Technology Campus, JD Block, Sector III, Bidhannagar, Kolkata, West Bengal 700098, India}
\address[univ2]{School of Computer Science and Software Engineering, University of Wollongong, New South Wales 2522,  Australia}

\begin{abstract}
i$^*$ models are inherently sequence agnostic. There is an immediate need to bridge the gap between such a sequence agnostic model and an industry implemented process modelling standard like \emph{Business Process Modelling Notation} (BPMN). This work is an attempt to build State Transition Models from i$^*$ models. In this paper, we first spell out the \emph{Naive Algorithm} formally, which is on the lines of Formal Tropos \cite{AFux}. We demonstrate how the growth of the State Transition Model Space can be mapped to the problem of finding the number of possible paths between the Least Upper Bound (\textit{LUB}) and the Greatest Lower Bound (\textit{GLB}) of a $k$-dimensional hypercube Lattice structure. We formally present the mathematics for doing a quantitative analysis of the space growth. The \emph{Naive Algorithm} has its main drawback in the \emph{hyperexponential explosion} caused in the State Transition Model space. This is identified and the \emph{Semantic Implosion Algorithm} is proposed which exploits the temporal information embedded within the i$^*$ model of an enterprise to reduce the rate of growth of the State Transition Model space. A comparative quantitative analysis between the two approaches concludes the superiority of the \emph{Semantic Implosion Algorithm}.
\end{abstract}

\begin{keyword}
\texttt{i$^*$, state transition model, model transformation}
\end{keyword}

\end{frontmatter}


\section{Introduction}
The use of models for software development has become a very standard software engineering practise. The advantages of using modelling notations to obtain the working view of a system or software, before actually coding it, is well-known. The most prominent benefits of modelling a system is to identify risks of failure before the coding of the system / software actually begins. Also, the use of standard modelling notations like UML helps in automated generation of code snippets.

A system or enterprise is modelled during the design phase of the development life cycle, only after the requirement specifications have been frozen by the consumer. Requirements Engineering helps in maintaining and documenting the user requirements. Requirement specifications are finalized only after multiple communications between the designer and the consumer. It is always beneficial to both the consumer and the designer if a working model of the system / enterprise can be obtained during the requirement analysis phase of development. \textit{i$^*$ models} were proposed keeping this in mind. The i$^*$ model provides an abstract sequence-agnostic view of the system to the consumer. In other words, it identifies actors and how they interact with each other. The i$^*$ model does not specify an activity work-flow or data-flow to the consumer. Temporal information is not specified anywhere within an i$^*$ model. This graphical representation of the system acts as a dashboard to the consumer where he can specify changes and be sure that the developer is in sync with what the consumer requires. 

Merely developing a formal modelling tool for requirement analysis only does not help the software engineering community much. i$^*$ models can have a huge impact on the development life cycle of systems / enterprises if we can map them to activity diagrams, and work-flow and data-flow models. Sequential or temporal characteristics are an inherent property of any standard business process model like BPMN or Petri-Nets. Without any control flow information, i$^*$ models prove to be futile. Again, since i$^*$ models are supposed to be sequence-agnostic, proposing modifications over i$^*$ models to incorporate temporal information, changes the very semantics and purpose of the i$^*$ modelling notation.The need of the hour is, thus, to bridge this gap between a sequence-agnostic requirements analysis model and a control-flow specific business process model. This paper is a conscious effort towards bridging this gap. 

The most obvious solution to this problem is using the brute-force method. Fuxman introduced the concept of actor instances and how dependencies, assertions, possibilities, and invariants can exist in either of three states - \textit{Not Created}, \textit{Created but Not Fulfilled}, and \textit{Fulfilled} \cite{AFux}. The \textit{Naive Algorithm} extends this concept to Goals, Tasks, Resources, and Dependencies existing within an i$^*$ model. It assumes that every model element goes through the above three states and makes two state transitions to reach the \textit{Fulfilled} state from the \textit{Not Created} state. Using this brute-force method to generate all possible state transition models corresponding to an i$^*$ model, results in an explosion within the state transition model space. This is identified in the following section. It is interesting to observe that, although an i$^*$ model is sequence agnostic, yet there exists some features or modelling constructs of the i$^*$ model that provide some temporal insight into the underlying system / enterprise. For instance, every dependency has a \emph{cause-effect property} in the sense that it is only when a dependee satisfies or fulfils a requirement of the depender does the dependency become satisfied. The \textit{Semantic Implosion Algorithm} identifies these untapped temporal characteristics and tries to contain the rate of growth of the state transition model space corresponding to an i$^*$ model. Simulation results reveal that the \textit{Semantic Implosion Algorithm} indeed outperforms the \textit{Naive algorithm} and provides a drastic improvement over the brute-force method. 

The rest of the paper is structured as follows. Section \ref{SOA} provides a review on existing techniques for transforming models. This section identifies that i$^*$ models have not beeen transformed to sequential models so far. The next section (Section \ref{Dev}) details out the \textit{Naive Algorithm} and the \textit{State Implosion Algorithm}. The drawbacks of the \textit{Naive Algorithm} are identified and the \textit{Semantic Implosion Algorithm} is proposed as a solution to these drawbacks. Section \ref{ER} performs a detailed simulation where both the algorithms are applied to the same classes of i$^*$ models and their behaviour are observed, compared and contrasted. Finally, section \ref{Concl} concludes the paper.
\section{State-of-the-Art}
\label{SOA}
Sendall and Kozaczynski had already identified Model Transformation as the central driving force behind Model-Driven Software Development \cite{SSWK}. Model Transformation represents the daunting challenge of converting higher-level abstraction models to platform-specific implementation models that may be used for automated code generation. Performing a model transformation requires a clear understanding of the abstract syntax and semantics of both the source and target.

Most Model-Driven Engineering practises offer a black box view of the transformation logic making it difficult to observe the operational semantics of a transformation. Most strategies work with lower levels of abstraction and encounter several limitations. In \cite{JSGK}, the authors propose a Domain Specific Language over Colored Petri-Nets - called Transformation Nets - that provides a high level of model transformation abstraction. An integrated view of places, transitions, and tokens, provide a clear insight into the previously hidden operational semantics.

Model transformation plays a vital role in bridging the gap between non-successive phases of the software development life cycle. \cite{SMAS} presents one such attempt to bridge the gap between system designers and system analysts. A model generated by the designer is transformed to a model suitable for conducting analysis. the outcome of the analysis is mapped back into the design domain. The authors work with \textit{UML2Alloy} - a tool that takes a UML Class diagram augmented with OCL constraints and converts it into the Alloy formal representation. Design inconsistency analysis is done on the Alloy representation. Alloy creates counter examples for any such inconsistency and converts it back into a UML Object diagram. This paper tries to do model transformation for bridging the gap between the Requirements phase and the Design phase of the development life cycle.

Creating a wide array of formal models for enhancing the system engineering process, proves to have time and cost overheads. Kerzhner and Paredis use model transformations to achieve this objective, overcoming the overheads, in \cite{AAK}. Formal models are used to specify the structures of varying design alternatives and design requirements, along with experiments that conform the two. These models are represented using the Object Management Group's Systems Modelling Language (OMG SysMLTM). Model transformation is then used to transform design structures into analysis models by combining the knowledge of reusable model libraries. Analysis models are transformed into executable simulations which help in identifying possible system alternatives. Model transformation plays a vital role in this work.

Mussbacher, \textit{et al}, have performed a detailed comparison of six different modelling approaches in \cite{GM}. The modelling approaches that were assessed include Aspect-oriented User Requirements Notation (\textit{AoURN}) \cite{GM2}, Activity Theory (\textit{AT}) \cite{GG}, The Cloud Component Approach (\textit{CCA}), Model Driven Service Engineering (\textit{MDSE}) \cite{SBK}, Object-oriented Software Product Line Modelling (\textit{OO-SPL}) \cite{ACBC}, and Reusable Aspect Models (\textit{RAM}) \cite{JKJK1, JKJK2}. The comparison criteria werre grouped into two broad categories - \textit{Modelling Dimensions} and \textit{Key Concepts}. Modelling dimensions include properties like Phase, Notation, and Units of Encapsulation. Key concepts, on the other hand, provide an insight into parameters like Paradigm, Modularity, Composability, Traceability and Trade-off Analysis. Of these six approaches, \textit{AoURN} \cite{GM2,GM3} and \textit{OO-SPL} \cite{ACBC} are of interest to this work, as both these approaches are applicable in the Early and Late Requirements phases of software development. the i$^*$ modelling notation belongs to this approach. In fact, \textit{AoURN} is based on the ITU-T Z.151 \cite{URN} standard that uses Goal-oriented Requirements Language (GRL), that is based on i$^*$ modelling. \textit{AoURN} is machine analysable and can perform scenario regression tests, goal-model evaluation, and ttrade-off analysis. Unlike the other modelling approaches, \textit{AoURN}  provides structural, behavioural, and intentional views, along with generic support for qualities and non-functional properties. It is purely graphical in nature.

The importance of i$^*$ modelling in Requirements Engineering has been well established in the last couple of years. Model transformations help in reducing the time and cost overheads associated with developing formal models for all possible design strategies across different architectures. There exists solutions to transform design models into execution sequences and perform various types of analyses. However, no work has been done so far on transforming requirement specification models to design models. Bridging this gap will help in identifying risks and failures during the Requirement Specification and Analysis phase of software development itself. Also, modern enterprises must ensure that they conform to a well-defined set of compliance rules involving government laws and regulations. Compliance checking deals with ensuring that an enterprise is system compliant. Although compliance rules can be defined as temporal properties on the system, compliance conformance cannot be verified with the i$^*$ model as it is typically sequence agnostic. In order to perform any type of model checking, we must first transform such a sequence-agnostic i$^*$ model into some form of state transition model that provides an insight into the possible sequence of activities within the enterprise. However, since it is an intuitive extraction of state transition models from an i$^*$ model, we might not be restricted to one particular unique solution. Rather, such a model transformation will be a one-to-many mapping. This work takes a leap in the efforts to bridge the gap between requirement models and design models. Two algorithms are presented and discussed that achieve this. A quantitative analysis is also performed between the two and the superiority of the \textit{Semantic Implosion Algorithm} over the \textit{Naive Algorithm} is established.

\section{Developing State Transition Models from an i* model}
\label{Dev}
The primary aim of this research is to analyze an i* model and develop all possible state transition models that can be derived from the given i* model. The challenge as well as motivation behind this work lies in the fact that i* models are sequence agnostic. However, without identifying a sequence of operations within the enterprise, it becomes very difficult to check and verify temporal properties and compliance rules within the system. Again, it is to be kept in mind that since an i* model is sequence agnostic, we cannot deterministically establish one single state transition model that corresponds to a given i* model. The output of this work will generate a set of state transition models, each of which satisfy the specification of the i* model. Once we obtain this valid set of plausible state transition models, we can apply some user defined enterprise specific compliance rules that fine tunes this set of probable state models. This final set of pruned state transition models can then be reverted back to the Enterprise owner in order to verify the requirements.

In this work we are considering the more detailed strategic relationship (SR) diagram of an i* model. The SR-diagram is much more comprehensive than its strategic dependency (SD) counterpart and encompasses all the dependency information that is captured in the SD-diagram. In fact, an SD-diagram represents the dependencies between different actors but does not exactly depict which particular model element of the depender is dependent on which particular model element of the dependee. The SR model is much more elaborate in this sense.

\subsection{The Naive Algorithm}
\label{sec:NA}
The simplest and most obvious solution to develop the set of all possible state transition models from an i* model, is to consider each model element separately and assume that they can exist in either of 3 possible states – \emph{Not Created (NC), Created Not Fulfilled (CNF),} and \emph{Fulfilled (F)} - and apply the \textit{brute-force} approach. In the first phase of this research, we assume a single instance of each model element appearing in the SR-diagram, i.e., each goal, task, or resource appearing in the SR-diagram represents a single instance of the corresponding model element. We obtain sequences of states or state transition models by evaluating the all possible permutations of the model elements and the state in which they exist.

\begin{figure}[h]
\centering
\includegraphics[width=0.8\textwidth, page=1]{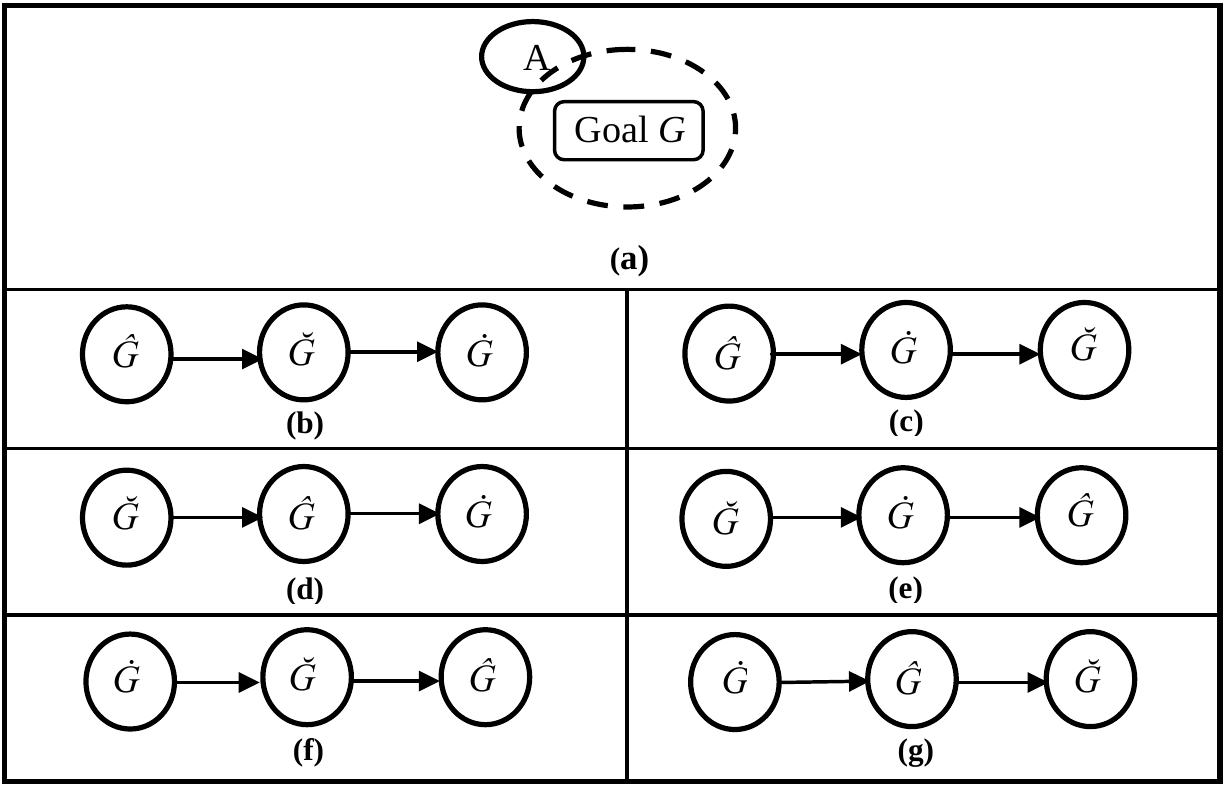}
\caption{\textbf{(a)} Actor A with a single goal G; \textbf{(b) – (g)} Possible state transition models that can be derived by permuting the state space}
\label{fig:oneGoal}
\end{figure}

Let us demonstrate the above concept with an example. Consider the simplest possible SR diagram with one actor consisting of only one goal G. This is shown in figure \ref{fig:oneGoal}.a. The goal G can be in either of three states – \textit{Not Created} denoted by \emph{$\hat{G}$}, \textit{Created Not Fulfilled} denoted by \emph{$\breve{G}$}, and \textit{Fulfilled} denoted by \emph{$\dot{G}$}. These three states give rise to 3! state transition models as shown in figures \ref{fig:oneGoal}.b to \ref{fig:oneGoal}.g.

However, out all these 6 state transition models, only figure 1.b is semantically correct. All the other state transition models are semantically inconsistent as a model element can go through its possible states in exactly one possible sequence – NC ($\hat{G}$) $\rightarrow$ CNF ($\breve{G}$) $\rightarrow$ F ($\dot{G}$). We call this sequence the \textit{default sequence}, and must be satisfied by all model elements. Now, let us increase the complexity by incorporating one more model element in the SR-diagram, i.e., let there exist two model elements in the SR-diagram. These two model elements can belong to the same actor or to two different actors. In either case, the complexity analysis remains the same.

\begin{figure}[h]
\centering
\includegraphics[width=0.8\textwidth, page=2]{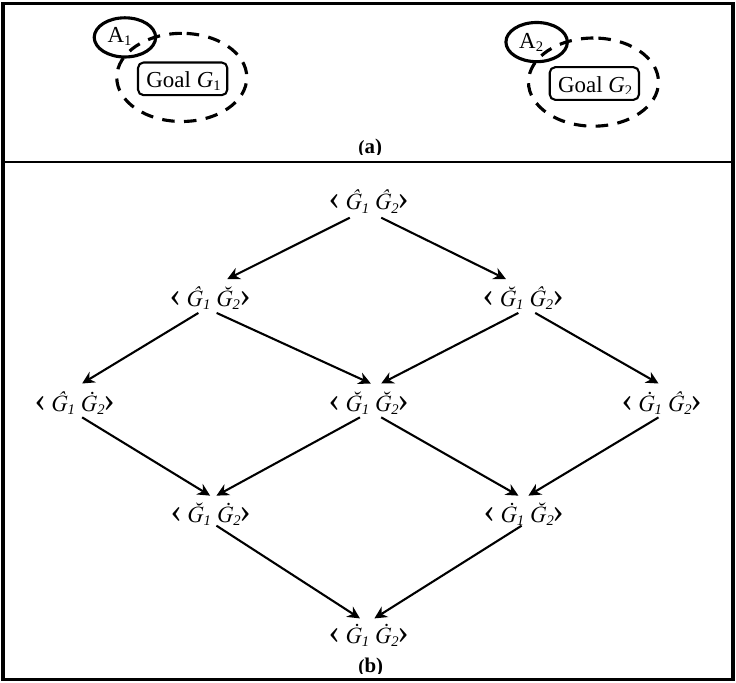}
\caption{\textbf{(a)} Actors A1 and A2 with goals G1 and G2, respectively; \textbf{(b)} The State Sequence Graph over the set of 3$^2$ = 9 possible states}
\label{fig:twoGoal}
\end{figure}

Let \emph{A$_1$} and \emph{A$_2$} be two different actors, each with a single goal node \emph{G$_1$} and \emph{G$_2$}, respectively. Since each goal can be in either of 3 states, the total number of possible combined states is $3^2$ (= 9). However, since both \emph{G$_1$} and \emph{G$_2$} must individually satisfy the \textit{default sequence}, it is interesting to observe the valid state transition sequences that do not violate the individual \textit{default sequences}. We draw a \textit{State Sequence Graph} that maps all the possible state transition paths from the source node – denoted by \emph{(${\hat{G}_1}$ ${\hat{G}_2}$)} – to the destination node – denoted by \emph{(${\dot{G}_1}$ ${\dot{G}_2}$)}. Figure \ref{fig:twoGoal}.b illustrates the \textit{State Sequence Graph} for two model elements.

The \textit{State Sequence Graph} has all the 9 possible combined state representations as vertices. These vertices are connected in the form of a mesh as all state transitions do not satisfy the \textit{default sequence}. Each path, in the \textit{State Sequence Graph}, from the source node \emph{(${\hat{G}_1}$ ${\hat{G}_2}$)} to the destination node \emph{(${\dot{G}_1}$ ${\dot{G}_2}$)} defines a semantically valid set of state transitions. In other words, each path represents a state transition model. Thus, with two model elements, we obtain 6 possible state transition models that satisfy the \textit{default sequences} of the individual model elements.
\\

\noindent\textbf{Definition:} \textit{State Sequence Graph}

A \emph{State Sequence Graph}, \emph{G},  can be defined as a 2-tuple $\langle$V,E$\rangle$  where \emph{V} represents the set of vertices and \emph{E} represents a set of directed edges such that,
\begin{enumerate}
  \item Each state ($\bar{G}_1$ $\bar{G}_2$.... $\bar{G}_n$) $\in$ V is an \emph{n}-tuple that represents each of the \emph{n} involved model elements in either of 3 possible states \emph{NC, CNF,} or \emph{F}, denoted by the generic symbol $\bar{G}_k$.
  \item Each directed edge \emph{e$_{ij}$} $\in$ E is directed from vertex \emph{v$_i$} to vertex \emph{v$_j$} such that \emph{v$_i$} $\rightarrow$ \emph{v$_j$} satisfies the \textit{default sequence} for any one of the \emph{n} model elements represented in every vertex notation. This implies that \emph{v$_i$} $\rightarrow$ \emph{v$_j$} represents either of the following –
      \begin{enumerate}
        \item Some goal $\bar{G}_i$ goes from the \emph{NC} state to the \emph{CNF} state, denoted by ($\bar{G}_1$... $\hat{G}_i$....$\bar{G}_n$) $\rightarrow$ ($\bar{G}_1$... $\breve{G}_i$....$\bar{G}_n$), or
        \item Some goal $\bar{G}_i$ goes from the \emph{CNF} state to the \emph{F} state, denoted by ($\bar{G}_1$... $\breve{G}_i$....$\bar{G}_n$) $\rightarrow$ ($\bar{G}_1$... $\dot{G}_i$....$\bar{G}_n$)
      \end{enumerate}
  \item The number of vertices in the vertex set V is 3$^n$, i.e., \textbar V\textbar = 3$^n$
  \item Each path from the source vertex ($\hat{G}_1$ $\hat{G}_2$ ... $\hat{G}_{n-1}$ $\hat{G}_n$) to the sink vertex ($\dot{G}_1$ $\dot{G}_2$ ... $\dot{G}_{n-1}$ $\dot{G}_n$) represents a valid state transition sequence that satisfies the \textit{default sequence} of each individual model element \emph{G$_1$, G$_2$, ..., G$_{n-1}$, G$_n$}, i.e., every unique path ($\hat{G}_1$ $\hat{G}_2$ ... $\hat{G}_{n-1}$ $\hat{G}_n$) $\rightarrow$ .... $\rightarrow$ ($\dot{G}_1$ $\dot{G}_2$ ... $\dot{G}_{n-1}$ $\dot{G}_n$) represents a state transition model
\end{enumerate}

\begin{figure}[h]
\centering
\includegraphics[width=0.9\textwidth, page=3]{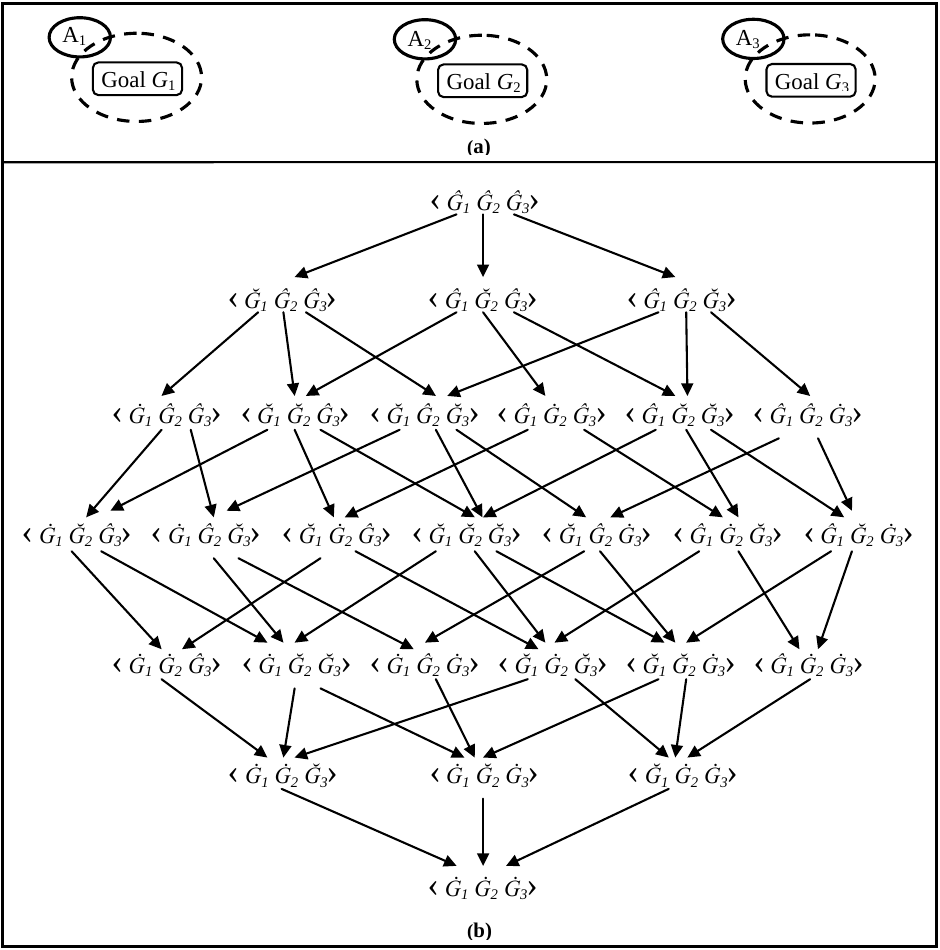}
\caption{\textbf{(a)} Actors \emph{A$_1$, A$_2$, and A$_3$} with goals \emph{G$_1$, G$_2$, and G$_3$}, respectively; (b) The State Sequence Graph over the set of 3$^3$ = 27 possible states}
\label{fig:threeGoal}
\end{figure}

The next level of complexity involves 3 different model elements. The analysis remains the same irrespective of how these 3 model elements are distributed between actors. Let \emph{G$_1$, G$_2$} and \emph{G$_3$} be the three different goals plotted in the SR-diagram. As mentioned above, since each goal can be in either of 3 states, this particular situation will result in a state space with 3$^3$(= 27) combined states. The \textit{State Sequence Graph} obtained is shown in Figure \ref{fig:threeGoal}.b.

A detailed reachability analysis using Depth-First Search yields 90 different paths that can be used to reach the sink vertex ($\dot{G}_1$ $\dot{G}_2$ $\dot{G}_3$) from the source vertex ($\hat{G}_1$ $\hat{G}_2$ $\hat{G}_3$). Each of these paths represents a plausible set of state transitions such that none of the 3 goals \emph{G$_1$, G$_2$, and G$_3$} violate the \textit{default sequence}. Thus, with 3 model elements in the SR-diagram we get 90 possible State Transition Models that correlate to the given i* model.

\subsubsection{Counting Multi-dimensional Lattice Paths}
In general, it is interesting to observe the number of paths  within a \textit{State Sequence Graph} corresponding to an i$^*$ model with $k$ model elements. It is intuitive from the above case studies that the state space grows exponentially as a function $f(k)=3^k$. This is because each of the model elements can exist in either of $3$ states. The growth function representing the growth of the state transition model space is far more complex. Before going into the details of an upper bound representing the growth of the state transition model space, we need to keep in mind that every model element is initially in the \textit{Not Created} state and it needs $2$ transitions to reach the \textit{Fulfilled} state. Thus, the distance covered by each model element is always $2$.

Consider the case where $k=2$. Since each model element needs to cover a distance of $2$, we can consider $(\hat{\emph{P$_1$}}\hat{\emph{P$_2$}})$ and $(\dot{\emph{P$_1$}}\dot{\emph{P$_1$}})$ as the \emph{Least Upper Bound} and the \emph{Greatest Lower Bound} of a $2\times2$ lattice. In general, the number of paths on a \emph{n$_1$}$\times$\emph{n$_2$} lattice is given by -
\begin{equation}
  \emph{L$_P$} = \binom{\emph{n$_1$}+\emph{n$_2$}}{\emph{n$_1$}} = \frac{(\emph{n$_1$}+\emph{n$_2$})!}{\emph{n$_1$}! \emph{n$_2$}!}
\end{equation}
So for a $2\times2$ lattice structure, we have -
\begin{equation}
  \emph{L$_P$} = \binom{2+2}{2} = \frac{(2+2)!}{2!\ 2!} = \frac{4!}{2!\ 2!} = \frac{24}{4} = 6. \nonumber
\end{equation}
This is exactly what we obtain from our empirical study in Figure \ref{fig:twoGoal}.

When $k=3$, we can represent the set of all possible transitions from $(\hat{\emph{P$_1$}}\hat{\emph{P$_2$}}\hat{\emph{P$_3$}})$ to $(\dot{\emph{P$_1$}}\dot{\emph{P$_1$}}\dot{\emph{P$_3$}})$ as $3$-dimensional cubic lattice. Again, since each model element makes $2$ transitions to be fulfilled, hence, we obtain a $2\times2\times2$ $3$-dimensional cubic lattice. In general, the number of paths in a $3$-dimensional cubic lattice with dimensions (\emph{n$_1$}, \emph{n$_2$}, \emph{n$_3$}) is given by -
\begin{equation}
  \emph{L$_P$} = \binom{\emph{n$_1$}+\emph{n$_2$}+\emph{n$_3$}}{\emph{n$_1$},\emph{n$_2$},\emph{n$_3$}} = \frac{(\emph{n$_1$}+\emph{n$_2$}+\emph{n$_3$})!}{\emph{n$_1$}! \emph{n$_2$}! \emph{n$_3$}!}
\end{equation}
So for a $3$-dimensional cubic lattice with dimensions(2, 2, 2), we have -
\begin{equation}
  \emph{L$_P$} = \binom{2+2+2}{2,2,2} = \frac{(2+2+2)!}{2!\ 2!\ 2!} = \frac{6!}{2!\ 2!\ 2!} = \frac{720}{8} = 90. \nonumber
\end{equation}
Again, this is exactly what we obtain from our empirical study in Figure \ref{fig:threeGoal}.

To generalize the upper bound on the growth function of the state transition model space, if we have a $k$-dimensional hypercube lattice with dimensions (\emph{n$_1$}, \emph{n$_2$}, ..., \emph{n$_k$}), then the number of paths is given by -
\begin{equation}
  \emph{L$_P$} = \binom{\emph{n$_1$}+\emph{n$_2$}+ ... +\emph{n$_k$}}{\emph{n$_1$},\emph{n$_2$},...,\emph{n$_k$}} = \frac{(\emph{n$_1$}+\emph{n$_2$}+...+\emph{n$_k$})!}{\emph{n$_1$}! \emph{n$_2$}!... \emph{n$_k$}!} = \frac{(\sum_{i=1}^k \emph{n$_i$})!}{\prod_{i=1}^k (\emph{n$_i$}!)}
\end{equation}
Irrespective of the number of model elements involved, since each model element travels a distance of 2 to become fulfilled, we have the condition $\forall_{i=1}^k$, \emph{n$_i$}=2. The total number of paths is given by -
\begin{equation}
  \label{EQ:Path}
  \emph{L$_P$} =\frac{(\sum_{i=1}^k 2)!}{\prod_{i=1}^k (2!)}= \frac{(2k)!}{2^k}.
\end{equation}

\begin{table}[t]
\caption{Rate of growth of space w.r.t. the number of model elements} 
\centering 
\begin{tabular}{c c c} 
\hline\hline 
No. of Process Elements & State Space & State Transition Model Space \\ [0.5ex] 
\hline 
5 & 243 & 113400\\
10 & 59049 & 2.37588E+15\\
15 & 14348907 & 8.09487E+27\\
20 & 3486784401 & 7.78117E+41\\
25 & 8.47289E+11 & 9.06411E+56\\
30 & 2.05891E+14 & 7.74952E+72\\
35 & 5.00315E+16 & 3.48622E+89\\
40 & 1.21577E+19 & 6.5092E+106\\
45 & 2.95431E+21 & 4.2227E+124\\
50 & 7.17898E+23 & 8.289E+142\\
55 & 1.74449E+26 & 4.4083E+161\\
60 & 4.23912E+28 & 5.8022E+180\\
65 & 1.03011E+31 & 1.7528E+200\\
70 & 2.50316E+33 & 1.1403E+220\\
75 & 6.08267E+35 & 1.5123E+240\\
80 & 1.47809E+38 & 3.8999E+260\\
85 & 3.59175E+40 & 1.876E+281\\
\hline 
\end{tabular}
\label{table:SpaceGrowth} 
\end{table}
Equation \ref{EQ:Path} can be used to generate a data set and observe how the state space and the state transition model space grows with increasing number of model elements in the i$^*$ model. Table \ref{table:SpaceGrowth} represents such a data set as the number of model elements increases from 5 to 85 in steps of 5. Data thus obtained can be plotted on a graph and the trends may be observed. Figure \ref{fig:NaiveGraph} below depicts the rate of growth for both the state space and the state transition model space with respect to the number of model elements depicted in the given i* model.

\begin{figure}[h]
\centering
\includegraphics[width=0.7\textwidth, page=4]{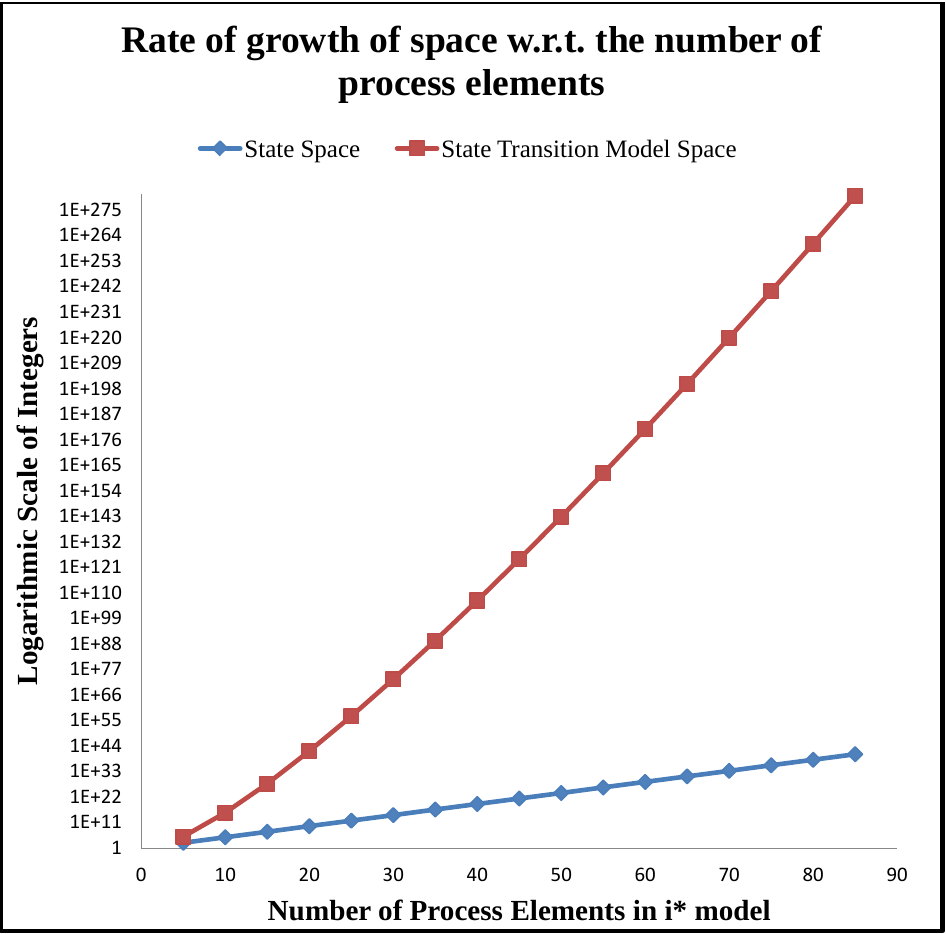}
\caption{Graph depicting the rate of growth of the state space and state transition model space with respect to the number of model elements in the i* model for the Naive Algorithm [\textit{To be reproduced in color on the Web and in black-and-white in print}]}
\label{fig:NaiveGraph}
\end{figure}

Interpretation of the graph is quite interesting. Some of the more interesting observations are as follows:
 \begin{enumerate}
   \item The reader should not to be misled by the \textit{linear} nature of the growth curves. A careful analysis of the graph reveals that the vertical axis represents a \textit{Logarithmic scale} where the values represent exponentially increasing integers. The values range from 1 to $1.876E+281$. Linear curves on a Logarithmic scale represent Exponential growth functions. In fact, the state space growth function, as represented by the blue line, actually represents the growth function $f(k)=3^k$. The growth function of the state transition model space, as represented by equation \ref{EQ:Path}, is represented by the red line.
   \item Another significant observation here is that the gradient of the blue line is much less than that of the red line. This implies that although both the state space and the state transition model space grows exponentially, the rate of growth for the state transition model space is much higher compared to that of the state space. In fact, the values in Table \ref{table:SpaceGrowth} reveal that, in every step, the state space grows by a factor of $10^2 - 10^3$, whereas the state transition model space grows by an approximate factor of $10^{19} - 10^{20}$. This is really huge in terms of the rate of growth.
 \end{enumerate}

  We can conclude from the above data that the Naive Algorithm causes a \textit{hyperexponential explosion} in the state transition model space. The growth curve of the state transition model space is so steep that it reaches infinitely large values for very small values of \emph{k}, the number of model elements in the i* model. It is evident from the nature of the curves that the state transition model space becomes quite unmanageable when we are looking at the i* model of an entire enterprise, comprising of hundreds of model elements. Thus, it becomes necessary to extract partial sequence information that remain embedded within an i* model and perform some pruning activities while the state transition model space is being generated.

\subsubsection{The Naive Algorithm:}

\noindent\textit{Input}: SR-diagram of the i* model of an enterprise

\noindent\textit{Output}: The set of plausible state transition models that can be derived from the given i* model

\noindent\textit{Data Structure}: A \textit{List} for each actor that stores model elements of the actor
\begin{list}{\textit{Step-\arabic{qcounter}}:~}{\usecounter{qcounter}}
\item Select the $i$-th model element \textit{P$_i$} from the \textit{List} of model elements for the actor \textit{A$_j$}.
\item Remove \textit{P$_i$} from the \textit{List}.
\item \textit{P$_i$} can make two transition from \emph{P$_i$}-\texttt{Not Created} to \emph{P$_i$}-\texttt{Created Not Fulfilled} and from \emph{P$_i$}-\texttt{Created Not Fulfilled} to \emph{P$_i$}-\texttt{Fulfilled}, in that order.
\item Generate all possible execution traces by interleaving the \textit{default sequences} of all model elements that have been removed from the \textit{List}, such that, the \textit{default sequence} of the individual model elements is satisfied.
\item Repeat \emph{Steps 1-4} for all model elements \textit{P$_i$} residing within the Actor boundary, i.e., while the \textit{List} is not empty.
\item Repeat \emph{Step 5} for all actors in the i$^*$ model.
\item Perform cartesian product between the sets of state transition models as obtained for individual actors, to generate the set of possible state transition models for the entire i$^*$ model. 
\item Stop.
\end{list}

\subsection{The Semantic Implosion (SI) Algorithm }

The motive here is to prevent the \textit{hyperexponential explosion} of the state transition model space that is caused by the \textit{Naive Algorithm}. Although the \textit{Naive Algorithm} generates all possible state transition models that can be derived from an i* model, some filtering can be done on this model space. The simplest means of doing this is to feed  each possible model being generated into some standard Model Verifier like \textit{NuSMV} and check the model against user-defined temporal compliance rules, specified using some standard temporal language like CTL or LTL. However, since this needs to be done on the entire state transition model space, the time complexity of the entire process becomes unmanageable even when machine-automated.

The desirable situation here is to prevent the \textit{hyperexponential explosion} from occurring in the first place. We propose the Semantic Implosion Algorithm, or \emph{SIA}, that tries to achieve this. \emph{SIA} is based on the underlying hypothesis that although an i$^*$ model is sequence agnostic, there exists some embedded temporal information that can be extracted and exploited to reduce the plausible space of state transition models. Temporal compliance rules may be defined that further reduce the number of coherent state transition models.

Every model element \emph{P$_i$} residing within the SR-diagram of an actor is uniquely identified using a system variable \emph{V$_i$}. Every system variable \emph{V$_i$} can have either of three values - 0, 1, or 2 - representing the conditions Not Created ($\hat{\emph{P$_i$}}$), Created Not Fulfilled $\breve{(\emph{P$_i$}}$), and Fulfilled ($\dot{\emph{P$_i$}}$), respectively. Every time a new model element \emph{P$_j$} is encountered, a corresponding system variable \emph{V$_j$} is created and initialized to 0 representing the Not Created situation. This is reflected in the state transition model of the enterprise with a transition from the current state to a new state where the corresponding system variable \emph{V$_j$} becomes a member of the state variables.

\begin{figure}[h]
\centering
\includegraphics[width=0.8\textwidth, page=7]{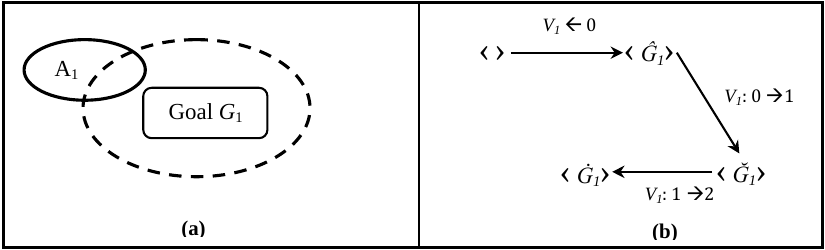}
\caption{\textbf{(a)} Actor \emph{A$_1$} with goal \emph{G$_1$}; (b) The corresponding State Transition Model}
\label{fig:SIASTM}
\end{figure}

The algorithm proceeds to explore the child model elements of a chosen parent model element. Before doing so, the corresponding system variable \emph{V$_j$} is updated to contain the value 1 and pushed onto a \textit{stack}. This is reflected in the state transition model with a state transition from the current state to a new state that reflects the fact that \emph{P$_j$} has been created but not fulfilled. A model element is said to be fulfilled when either it has no child model element (we have reached the actor boundary) or all its child model elements have been individually fulfilled. When this happens, the system variable \emph{V$_j$} corresponding to the parent model element \emph{P$_j$} is popped from the stack and updated with the value 2. A corresponding state transition is incorporated in the state transition model that reflects the fact that model element \emph{P$_j$} has been fulfilled. Figure \ref{fig:SIASTM} illustrates the state transition model corresponding to a single model element and how the corresponding system variable is incorporated and updated along each transition.

However, it is interesting to note how the child model elements of a particular parent are processed. The processing differs for \textit{task decompositions} and \textit{means-end decompositions}. A \textit{task decomposition} is an AND-decomposition and demands that all the child model elements be fulfilled in order to declare that the parent has also been fulfilled. A \textit{means-end decomposition}, on the other hand, is an OR-decomposition and provides alternate strategies to fulfill the parent model element. Let us elaborate on the consequences of these two decompositions.

\begin{figure}[h]
\centering
\includegraphics[width=0.6\textwidth, page=5]{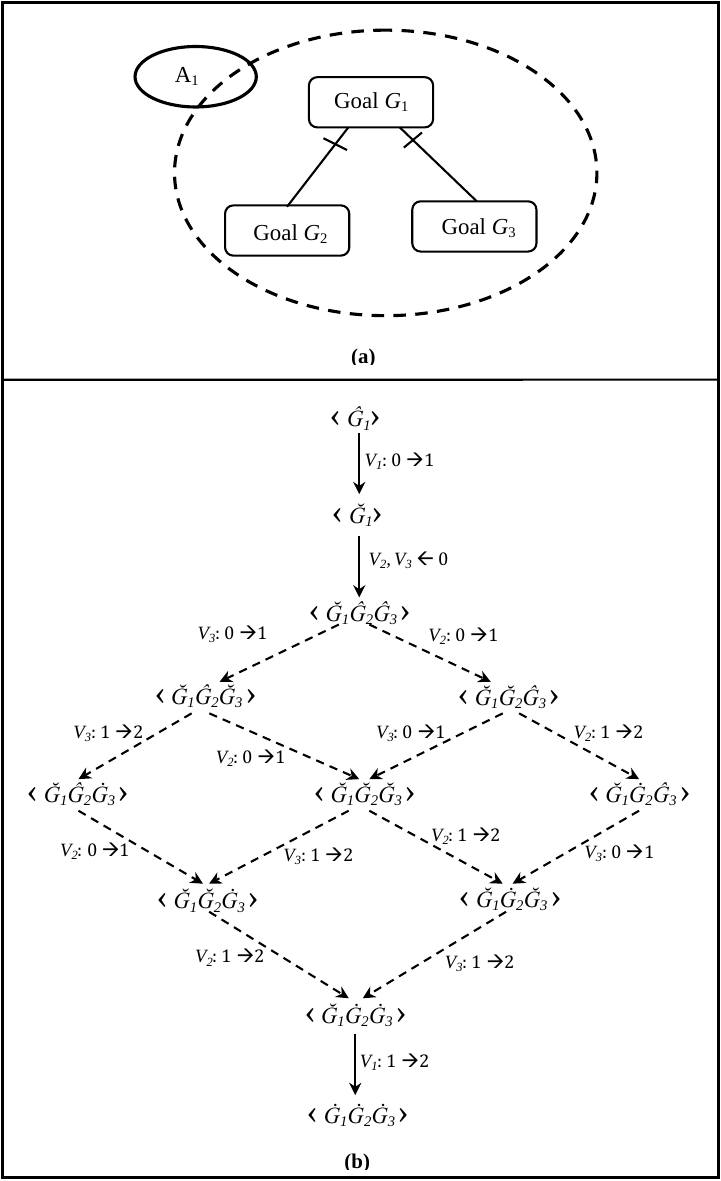}
\caption{\textbf{(a)} Actor \emph{A$_1$} with goals \emph{G$_1$, G$_2$, and G$_3$} connected through a task decomposition; (b) The corresponding set of all possible State Transition Models}
\label{fig:taskdec}
\end{figure}

A \textit{task decomposition} requires that all the child model elements be fulfilled before changing the state of the parent model element to the fulfilled state. However, since an i$^*$ model is sequence agnostic, the child model elements may be fulfilled in any random permutation. System variables associated with the child model elements should not defy the \textit{default sequence} defined in section \ref{sec:NA}. Let a model element \emph{P$_j$} be decomposed by a \textit{task decomposition} to a set of model elements $\langle\emph{P$_1$}, \emph{P$_2$}, ..., \emph{P$_m$}\rangle$. The system variables associated with these model elements are \emph{V$_1$}, \emph{V$_2$}, ..., \emph{V$_m$}, respectively. We define a state transition from the current state with \emph{V$_j$}=1 to a new state with the state variables \emph{V$_j$}=1, $\forall_{r=1}^m,\emph{V$_r$}$=0. There exists several execution permutations of the decomposed model elements that results in a state with the state variables, \emph{V$_j$}=1, $\forall_{r=1}^m,\emph{V$_r$}$=2. The set of all possible execution sequences can be defined using a lattice structure, similar to the ones shown in figures \ref{fig:twoGoal}, and \ref{fig:threeGoal}. Since all child model elements are fulfilled in this state (the \emph{GLB} of the lattice), we define another state transition in the state transition model that reflects the fact that the parent model element is also fulfilled, i.e., the new state has state variables \emph{V$_j$}=2, $\forall_{r=1}^m,\emph{V$_r$}$=2. The state transition model corresponding to such a \textit{task decomposition} is shown in figure \ref{fig:taskdec}.

The interpretation of the figure is quite interesting. The lattice structure represents the set of all possible execution sequences that result in the successful fulfillment of the task decomposition. As seen in section \ref{sec:NA}, the number of paths in a lattice structure for two model elements is 6. All of these 6 paths represent valid execution sequences or state transitions. Each path gives rise to a different state transition model. This implies that the task decomposition shown in figure \ref{fig:taskdec} gives rise to 6 possible state transition models. The \textit{Naive Algorithm}, on the other hand, would generate a lattice structure with three model elements and the number of possible state transition models would become 90. This is a significant reduction in the state transition model space. In fact, the significant observation here is that a lattice structure will be generated only where AND-decompositions take place. In other words, only AND-decompositions will increase the size of the state transition model space.

\begin{figure}[h]
\centering
\includegraphics[width=0.6\textwidth, page=6]{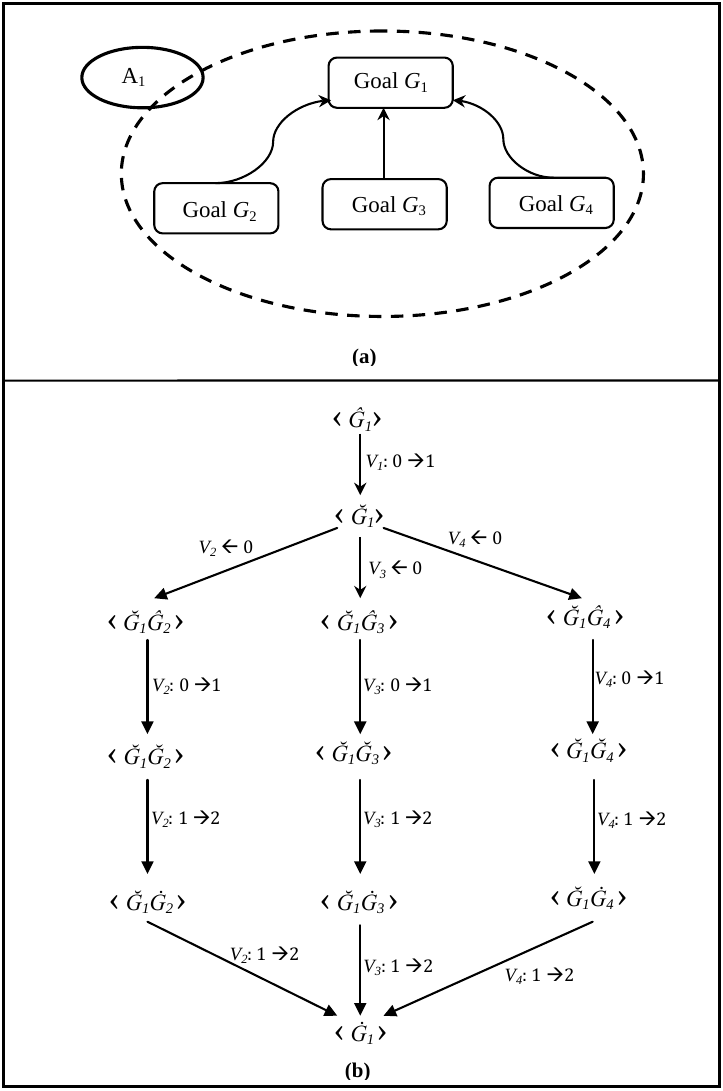}
\caption{\textbf{(a)} Actor \emph{A$_1$} with goals \emph{G$_1$, G$_2$, G$_3$, and G$_4$} connected through a means-end decomposition; (b) The corresponding State Transition Model}
\label{fig:MEdec}
\end{figure}

A \textit{means-end} decomposition is easier to handle. OR-decompositions, in general, do not increase the size of the state transition model space. Rather, if a particular model element \emph{P$_j$} decomposes via a \textit{means-end decomposition} into $k$ model elements $\langle\emph{P$_1$}, \emph{P$_2$}, ..., \emph{P$_k$}\rangle$, then we introduce $k$ different transitions from the current state (\emph{V$_j$}=1) to $k$ unique new states, each representing one of the $k$ alternate means (\emph{V$_j$}=1, \emph{V$_p$}=0, $\forall_{p=1}^k$). An OR-decomposition is characterized by the fact that fulfilling any one of the alternate means implies fulfilling the parent model element. Thus, each of these $k$ new states will make two transitions (labelled by \emph{V$_p$}:0$\rightarrow$1 and \emph{V$_p$}:1$\rightarrow$2, $\forall_{p=1}^k$) to reach their respective fulfillment states. Each alternate means will have a separate fulfillment state labelled by \emph{V$_j$}=1, \emph{V$_p$}=2, $\forall_{p=1}^k$. All the $k$ fulfillment states will converge to a final state that represents the fulfillment of the parent model element \emph{P$_j$} and is labelled by \emph{V$_j$}=2, $\vee_{p=1}^k\emph{V$_p$}$=2. The structure obtained is similar to the longitudinal lines on the globe of the earth. Figure \ref{fig:MEdec} illustrates this further.

\subsubsection{Some interesting Observations}
\begin{enumerate}
  \item Decompositions can be \emph{nested}. This implies that decompositions can occur within other decompositions. One particular decomposition link may be further blown up with a second decomposition. For instance, \textit{means-end decompositions} may be followed by a \textit{task decomposition} along one means-end link and a \textit{means-end decomposition} along some other means-end link. Figure \ref{fig:Nesteddec} illustrates this scenario. This nesting of decompositions does not require any modifications on the algorithm. The corresponding state transition models are built accordingly where the state transition sub-model of the nested decomposition is mereologically connected to the state transition model of the outer level decomposition.
      \begin{figure}[h]
        \centering
        \includegraphics[width=0.6\textwidth, page=8]{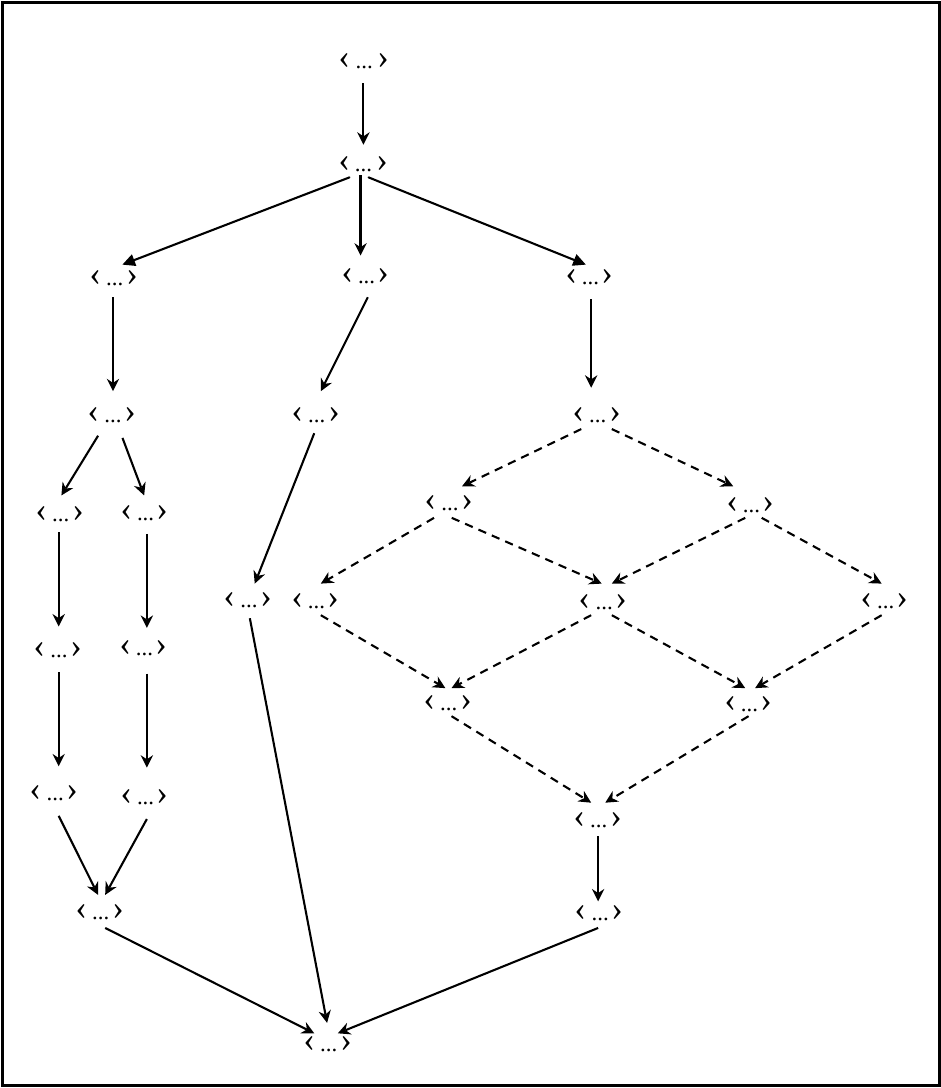}
        \caption{The State Transition Model corresponding to a nested decomposition. An outer mean-end decomposition contains another means-end decomposition along the leftmost means and a task decomposition along the rightmost means.}
        \label{fig:Nesteddec}
      \end{figure}
  \item It is interesting to note what happens if we reach a model element \emph{P$_i$}, located at the actor boundary of actor \emph{A$_i$}, that is dependent on some model element \emph{P$_j$} that is located at the actor boundary of actor \emph{A$_j$}. In that case, we assume that the model element \emph{P$_i$} will be fulfilled by \emph{A$_j$}, pop out the system variable \emph{V$_i$} from the \emph{stack} and set its value to 2. At the same time we introduce a  \emph{temporary transition} in the corresponding state transition model that changes the state of \emph{V$_i$} from Created Not Fulfilled(\emph{CNF}) to Fulfilled(\emph{F}). This is necessary as we cannot proceed with the construction of the state transition model of individual actors without this assumption. However, we need to maintain a list of all such dependencies. A \textit{Global List} is maintained that stores 2-tuples of the form $\langle\emph{depender variable, dependee variable}\rangle$. Once the state transition models of the individual actors have been built, the elements of the \textit{Global List} are accessed. Each element represents a dependency of the form $\langle\emph{\emph{V$_{ik}$}, \emph{V$_{jl}$}}\rangle$ and is interpreted as model element \emph{P$_k$} within actor \emph{A$_i$} depending on actor \emph{A$_j$} for model element \emph{P$_l$}. The \emph{temporary transition} in the state transition model of actor \emph{A$_i$} representing the change \emph{V$_k$}: $1\rightarrow2$ is replaced by two new transitions that connect the state transition models of actors \emph{A$_i$} (STM$_i$) and \emph{A$_j$} (STM$_j$). The first transition is established from the state in \emph{STM$_i$} having label \emph{V$_k$}=1 to the state in \emph{STM$_j$} having label \emph{V$_l$}=2. The second transition is placed from the state in \emph{STM$_j$} having label \emph{V$_l$}=2 to the state in \emph{STM$_i$} having label \emph{V$_k$}=2. $\langle\emph{\emph{V$_{ik}$}, \emph{V$_{jl}$}}\rangle$ is removed from the \textit{Global List}. Figure \ref{fig:depSTM} further illustrates this process.
      \begin{figure}[h]
        \centering
        \includegraphics[width=0.8\textwidth, page=9]{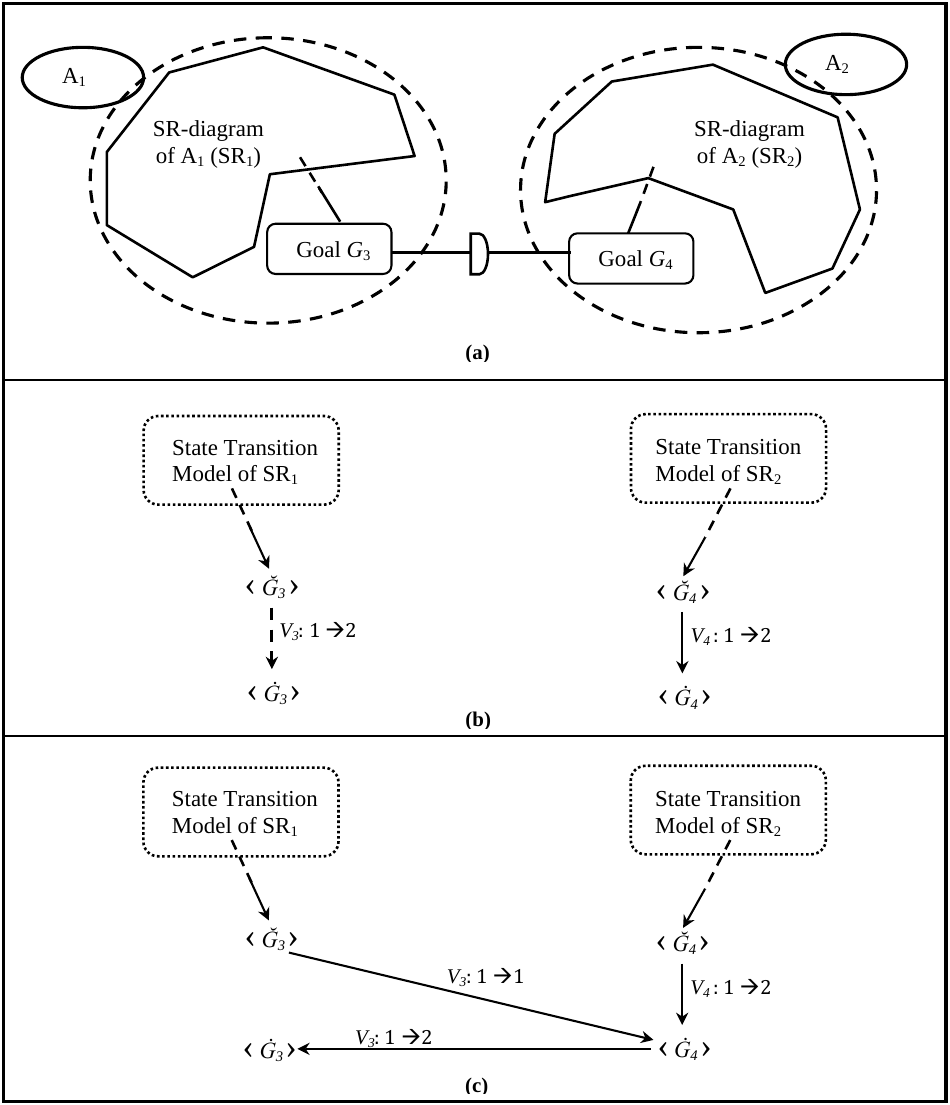}
        \caption{\textbf{(a)} Goal \emph{G$_3$} of actor \emph{A$_1$} dependant on Goal \emph{G$_4$} of actor \emph{A$_2$}; \textbf{(b)} Temporary transition from \emph{$\breve{G_3}$} to \emph{$\dot{G_3}$} introduced; \textbf{(c)}Resolution of the dependency by replacing the temporary transition with two permanent transitions}
        \label{fig:depSTM}
      \end{figure}
      \item Dependency resolution causes state transitions to be set up between states belonging to the state transition models of the \emph{depender} and the \emph{dependee}. If the \emph{depender} and \emph{dependee} have $M$ and $N$ possible state transition models, respectively, then we get a maximum of $M \times N$ combinations for interlinking the state transition models of the \emph{depender} and \emph{dependee}. Every dependency resolution must take place simultaneously in all the $M \times N$ combinations.
\end{enumerate}

 Let $n$ be the total number of model elements occurring in the SR-diagram of the enterprise. The terminating condition of the \textit{SI Algorithm} is given by the constraint, $\forall_{j=1}^n, \emph{V$_j$}$=2 and the \textit{Global Dependency List} is empty. The algorithm initiates with the model elements at the actor boundaries that do not stem from a parent model element. State transitions are defined in the corresponding state transition model as and when model elements are discovered, explored and fulfilled. Let us look into the \textit{Semantic Implosion Algorithm} now.


\subsubsection{The Semantic Implosion Algorithm:}

\noindent\textit{Input}: SR-diagram of the i* model of an enterprise

\noindent\textit{Output}: The set of plausible state transition models that can be derived from the given i* model

\noindent\textit{Data Structure}: A \textit{Local Stack} for each actor that stores model elements of the actor and a \emph{Global List} to keep track of dependencies between actors
\begin{list}{\textit{Step-\arabic{qcounter}}:~}{\usecounter{qcounter}}
    \item For every model element \emph{P$_i$} that is not at the end of a \textit{task decomposition} or \textit{means-end} link, assign a system variable \emph{V$_i$}=0. Perform a \textit{Depth-First Scan} of the SR-diagram of each actor starting at these boundary model elements.
    \item For any model element \emph{P$_j$} with \emph{V$_j$}=0, set \emph{V$_j$}=1 and push it onto the \textit{Local Stack}. Reflect this transition in the state transition model by plotting a transition from the \textit{Not Created} state to the \textit{Created Not Fulfilled} state. Label this transition \emph{V$_j$}:0$\rightarrow$1.
    \item Discover all model elements $\langle\emph{P$_1$}, \emph{P$_2$}, ..., \emph{P$_q$}\rangle$ that stem from the element \emph{P$_j$} and are connected to \emph{P$_j$} with \textit{task decomposition} or \textit{means-end} links. For each such element \emph{P$_k$}, initialize a system variable \emph{V$_k$} such that $\forall_{k=1}^q$\emph{V$_k$}=0.
    \begin{list}{\textit{\alph{acounter}})~}{\usecounter{acounter}}
      \item If \emph{P$_j$} is at an actor boundary with no elements stemming from it and with no dependencies to other actors, pop \emph{V$_j$} from the \emph{Stack} and set \emph{V$_j$}=2. Set up a corresponding transition in the state transition model from the \textit{Created Not Fulfilled} state to the \textit{Fulfilled} state. Label this transition \emph{V$_j$}:1$\rightarrow$2.
      \item If \emph{P$_j$} is dependent on some other actor for fulfillment, then pop \emph{V$_j$} and insert it into the \emph{Global List} with value \emph{V$_j$}=2. Insert a \textit{temporary transition} between states \textit{Created Not Fulfilled} and \textit{Fulfilled}. No need to label this transition as it is a \textit{temporary transition}.
      \item If \emph{P$_j$} undergoes a \emph{task decomposition} then we obtain several different state transition sub-models for the \emph{task decomposition} by permuting the order of execution of the decomposed model elements. Each such permutation can be considered to be a valid state transition sub-model and can be attached to the overall state transition model to obtain a set of unique state transition models for the actor.
      \item If \emph{P$_j$} undergoes a \emph{means-end decomposition} then we obtain multiple transitions from the current node in the same state transition model. Each transition represents an alternate strategy and is triggered by the corresponding guard condition. All the alternate state transitions emanating from the parent model element must converge at a state that represents that the parent model element has been fulfilled.
    \end{list}
    \item Repeat \emph{Steps 2-3} for all siblings of \emph{P$_j$} in all the state transition models generated for actor \emph{A$_i$}.
    \item Repeat \emph{Step 4}until the \emph{Local Stack} is empty. This leaves us with the set of plausible state transition models of an actor \emph{A$_i$}.
    \item Repeat \emph{Steps 1-5} to extract all the possible state transition models of all the actors in the i* model.
    \item Remove elements of the form $\langle\emph{\emph{V$_{ik}$}, \emph{V$_{jl}$}}\rangle$ from the \emph{Global List}.
    \item Remove the \textit{temporary transitions} corresponding to this dependency from all state transition models of actor \emph{A$_i$}.
    \item Insert transitions from the \emph{P$_k$}-\texttt{Created Not Fulfilled} state in all state transition models of actor \emph{A$_i$} to the \emph{P$_l$}-\texttt{Fulfilled} state in all state transition models of actor \emph{A$_j$}. Label these transitions \emph{V$_k$}:1$\rightarrow$1.
    \item Insert another set of transitions from the \emph{P$_l$}-\texttt{Fulfilled} state to the the \emph{P$_k$}-\texttt{Fulfilled} state in between all possible state transition models of actors \emph{A$_i$} and \emph{A$_j$}. Label these transitions \emph{V$_k$}:1$\rightarrow$2.
    \item Repeat \emph{Steps 7-10} until the \emph{Global List} is empty and all the dependencies have been resolved.
    \item Stop.
\end{list}

\section{Experimental Results}
\label{ER}
Let us perform some analytics on comparing and contrasting the behavior of the \textit{Naive Algorithm} and the \textit{Semantic Implosion Algorithm}. The two metrics that are used for this analysis are the \textit{State Space Size}(SSS) and the \textit{State Transition Model Space Size}(STMSS). However, since both algorithms share the concept of every model element going through $3$ states, the SSS metric will be the same for both algorithms and is defined by \textit{f(k)=3$^k$}. The STMSS metric is far more crucial in contrasting the behavioral differences between the two algorithms.

Figure \ref{fig:NaiveGraph} clearly illustrates the \textit{hyper-exponential explosion} caused by the \textit{Naive Algorithm} in the state transition model space. This is mainly due to the fact that the \textit{Naive Algorithm} considers all possible orderings of the model elements ensuring the \textit{default sequence} of each individual model element. A careful understanding of the \textit{SI Algorithm} reveals that, while the state transition model corresponding to every actor is being built, the state transition model space increases only when the following conditions hold:
\begin{enumerate}
  \item Whenever a \textit{Task Decomposition} is encountered. Suppose a task is decomposed to $k$ different model elements. Since an i$^*$ model is sequence agnostic, these $k$ model elements can be executed in any order. The set of all possible execution traces is given by a $k$-dimensional hypercube lattice with each dimension having distance $2$. As discussed in section \ref{sec:NA}, the state transition model space increases by a factor of $\frac{(2k)!}{2^k}$ as given by equation \ref{EQ:Path}. This implies that if the state transition model space representing the set of all possible state transition models already has $p$ models, a task decomposition into $q$ model elements causes the number of state transition models to become p\textbf{.}$\frac{(2q)!}{2^q}$. In general, if the SR-diagram of an actor within the i* model has $d$ task decompositions, and the number of possible alternate execution sequences generated by each of these task decompositions be given by \emph{\#Seq$_1$}, \emph{\#Seq$_2$}, ..., \emph{\#Seq$_d$}, then  state transition model space size is given by the following relation:
      \begin{equation}
      S = \prod_{i=1}^d \emph{\#Seq$_i$}
      \end{equation}
  \item Whenever a dependency is being resolved. \textit{Dependency resolution} needs to be done individually for every pair of models that can be extracted from the state transition model space of the \textit{depender} and the \textit{dependee}. If the state transition model spaces of actors \emph{A$_i$} and \emph{A$_j$} contain $M$ and $N$ models respectively, then irrespective of the number of dependencies between \emph{A$_i$} and \emph{A$_j$}, the state transition model space changes from $M+N$ to $M \times N$. Again, if actor \emph{A$_j$} requires dependency resolution with actor \emph{A$_k$}, and actor \emph{A$_k$} has $L$ state transition models, then the combined state transition model space has size $L \times M \times N$.
  \item Even if 2 actors \emph{A$_r$} and \emph{A$_s$} are not dependent on each other, \textit{Completeness} demands that the state transition model space of the entire i$^*$ model considers all possible combinations of the state transition models of actors \emph{A$_r$} and \emph{A$_s$}. Let \{\emph{S$_1$}, \emph{S$_2$}, ..., \emph{S$_n$}\} be the state transition model space sizes of the $n$ actors participating in an i$^*$ model. Then the size of the state transition model space of the entire enterprise ($S$) is given by the following relation:
      \begin{equation}
      \label{EQ:EnterpiseSize}
      S = \prod_{i=1}^n \emph{S$_i$}
      \end{equation}
\end{enumerate}

\textit{Dependency Resolution} and \textit{Completeness} conditions are both represented using the Cartesian Product relation. So, behaviour analysis boils to two basic steps. The first steps involves observing the growth of the state transition model space for each individual actor. Once this has been done for all the actors, the state transition model space for the entire enterprise is constructed.

\subsection{Actor Internal Analytics}
It is very difficult to predict the distribution of model elements of an i$^*$ model within the SR-diagrams of individual actors. Since this is the first step of behaviour analysis, we are concerned with the state transition model space growth of individual actors within an i* model. In order to generate a consistent data set, we assume a uniform distribution of model elements. We increase the number of model elements occurring within the SR-diagram of an actor in the i$^*$ model in steps of 5. Without loss of \textit{uniformity}, we assume that for every $5$ model element within an actor, there exists a task decomposition of $4$ elements.

We know that the \textit{Naive Algorithm} causes the state transition model space to grow according to equation \ref{EQ:Path}, i.e., STMSS$_N$ = $\frac{(2k_1)!}{2^{k_1}}$, where \emph{k$_1$} is the number of  model elements in the i$^*$ model. The \textit{Semantic Implosion Algorithm} grows only on the basis of task decompositions. The number of possible execution sequences generated by a 4-element task decomposition is obtained by substituting $k=4$ in equation \ref{EQ:Path}, i.e., $\frac{(2.4)!}{2^4}$ = $\frac{8!}{16}$ = $2520$. Since every 4-element task decomposition increases the state transition model space size by a factor of 2520, applying the Cartesian Product relation, we obtain the growth function of the \textit{SI Algorithm} to be given by STMSS$_S$ = $2520^{k_2}$, where \emph{k$_2$} is the number of 4-element task decompositions occurring within the SR-diagram of an actor. Table \ref{table:AIAnalytics} reflects such a data set.

\begin{table}[t]
\centering
\caption{Actor Internal Analytics}
\begin{tabular}{cccc}
\hline \hline
\multirow{2}{*}{\begin{tabular}[c]{@{}c@{}}No. of Process \\ Elements (\emph{k$_1$})\end{tabular}} & \multirow{2}{*}{\begin{tabular}[c]{@{}c@{}}No. of 4-element Task \\ Decompositions (\emph{k$_2$})\end{tabular}} & \multicolumn{1}{l}{Naive Algorithm} & \multicolumn{1}{l}{SI Algorithm} \\ \cline{3-4}
                                                                                         &                                                                                             & \multicolumn{1}{l}{STMSS$_N$ = $\frac{(2\emph{k$_1$})!}{2^\emph{k$_1$}}$}      & \multicolumn{1}{l}{STMSS$_S$ = $2520^\emph{k$_2$}$}   \\ \hline
5 & 1  & 113400 & 2520                             \\
10 & 2 & 2.37588E+15 & 6350400                          \\
15 & 3 & 8.09487E+27 & 1.6E+10                          \\
20 & 4 & 7.78117E+41 & 4.03E+13                         \\
25 & 5 & 9.06411E+56 & 1.02E+17                         \\
30 & 6 & 7.74952E+72 & 2.56E+20                         \\
35 & 7 & 3.48622E+89 & 6.45E+23                         \\
40 & 8 & 6.5092E+106 & 1.63E+27                         \\
45 & 9 & 4.2227E+124 & 4.1E+30                          \\
50 & 10 & 8.289E+142 & 1.03E+34                         \\
55 & 11 & 4.4083E+161 & 2.6E+37                          \\
60 & 12 & 5.8022E+180 & 6.56E+40                         \\
65 & 13 & 1.7528E+200 & 1.65E+44                         \\
70 & 14 & 1.1403E+220 & 4.16E+47                         \\
75 & 15 & 1.5123E+240 & 1.05E+51                         \\
80 & 16 & 3.8999E+260 & 2.64E+54                         \\
85 & 17 & 1.876E+281  & 6.67E+57                         \\ \hline
\end{tabular}
\label{table:AIAnalytics}
\end{table}

\begin{figure}[h]
\centering
\includegraphics[width=0.7\textwidth, page=10]{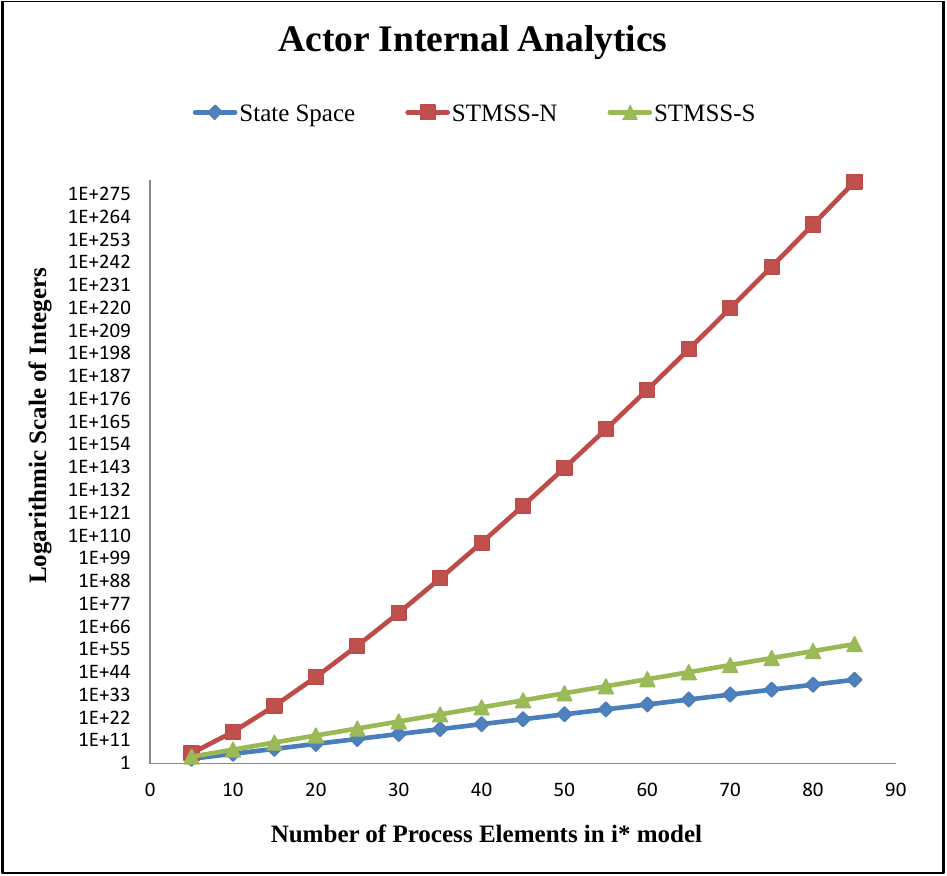}
\caption{Behaviour analysis with respect to the state transition model space of individual actors for the \textit{Naive Algorithm} (\texttt{STMSS-N}) and the \textit{Semantic Implosion Algorithm} (\texttt{STMSS-S}) as the number of model elements in the i$^*$ model varies [\textit{To be reproduced in color on the Web and in black-and-white in print}]}
\label{fig:AIAnalytics}
\end{figure}

The graph plotted on the basis of this data is shown in Figure \ref{fig:AIAnalytics}. It is interesting to analyze the graph. The vertical axis is again a \textit{Logarithmic Scale of Integers}. Straight lines in this plot represent exponential functions. The slopes of the straight lines are directly proportional to the rate of growth of the corresponding exponential function, i.e., greater the slope, the greater is the exponential rate of growth. The following observations can be concluded from the graph:
\begin{enumerate}
  \item The blue line depicts the growth of the state space and is consistent for both scenarios, given by $3^k$. As both algorithms have the underlying basis that every model element goes through 3 states, the state space growth remains the same.
  \item The green line depicts the behaviour of the \textit{Semantic Implosion Algorithm}. It is to be noted that the blue and green lines are very close to one another. This implies that the exponential rate of growth of the state transition model space as governed by the \textit{SI Algorithm} is almost the same as the exponential rate of growth of the state space.
  \item The red line depicts the exponential rate of growth for the \textit{Naive Algorithm}. The slope of the red line is much greater than those of the green and blue lines. This represents the \textit{hyper-exponential explosion} that is a characteristic of the \textit{Naive Algorithm}.
  \item A closer look at the STMSS values in Table \ref{table:AIAnalytics} reveals the fact that the STMSS metric increases by a factor of $10^{19}$ - $10^{20}$ for the \textit{Naive Algorithm} whereas for the \textit{SI Algorithm} the STMSS metric increases by a factor of $10^3$.
\end{enumerate}

The conclusions from Table \ref{table:AIAnalytics} and Figure \ref{fig:AIAnalytics} clearly indicate that the \textit{Semantic Implosion Algorithm} provides a huge improvement in the rate of growth of the state transition model space with respect to individual actors in comparison to the \textit{Naive Algorithm}.

\subsection{Inter-Actor Analytics}
\textit{Actor Internal Analytics} explore the growth of the state transition model space for each individual actor. \textit{Inter-Actor Analytics} provides an insight into how \textit{Actor Internal Analytics} impact the state transition model space growth rate of the entire i$^*$ model representing an enterprise. There are two events that impact \textit{Inter Actor Analytics} as follows:
\begin{enumerate}
	\item \textit{Density of Actors} participating in the i$^*$ model, and
	\item \textit{Distribution of Process Elements} within the actors.
\end{enumerate}
Let us individually analyse how these two parameters effect the growth rate of the state transition model space.

\subsubsection{Variation of Actor Density}
Let there be $n$ actors participating in an i$^*$ model. Let the size of the state transition model spaces of the individual actors be given by \textit{S$_{1}$, S$_{2}$, ..., S$_{n}$,} respectively. Assuming a uniform density of 5 model elements within individual actors, we try to evaluate the rate of growth of the state transition model space. Similar to the data in table \ref{table:SpaceGrowth}, we assume that every actor has a 4-element task decomposition.

The \textit{Naive Algorithm} affects the state transition model space by causing the space size to grow according to equation \ref{EQ:Path}. Replacing $k=5$, we get the state transition model space size of every actor as -
\begin{equation}
\emph{L$_P$} = \frac{(2.5)!}{2^5} = \frac{10!}{32} = 113400. \nonumber
\end{equation}

Since all the $n$ actors of the i$^*$ model have uniform distribution of model elements, the state transition model space size remains the same for all actors as given by equation \ref{EQ:Path}, i.e., $\forall_{i=1}^n$, S$_i$ = $\frac{(2k)!}{2^k}$. Combining equations \ref{EQ:Path} and \ref{EQ:EnterpiseSize}, we get the state transition model space size for the entire enterprise (S$_N$) as -
\begin{equation}
\emph{S$_N$} = (\frac{(2k)!}{2^k})^n
\end{equation}

The \textit{Semantic Implosion Algorithm}, on the other hand, causes the state transition model space of individual actors to grow only when task decompositions are encountered. Since we assume a 4-element task decomposition to exist in each actor, the state transition model space size of all the actors remains constant and is given by replacing $k=4$ in equation \ref{EQ:Path}.
\begin{equation}
\emph{L$_P$} = \frac{(2.4)!}{2^4} = \frac{8!}{16} = 2520. \nonumber
\end{equation}

Since $\forall_{i=1}^n$, S$_i$ = 2520, replacing this value in equation \ref{EQ:EnterpiseSize} the state transition model space size for the entire enterprise (S$_S$), as given by the \textit{SI Algorithm}, is -
\begin{equation}
\emph{S$_S$} = (2520)^n
\end{equation}

\begin{figure}[h]
	\centering
	\includegraphics[width=0.7\textwidth, page=12]{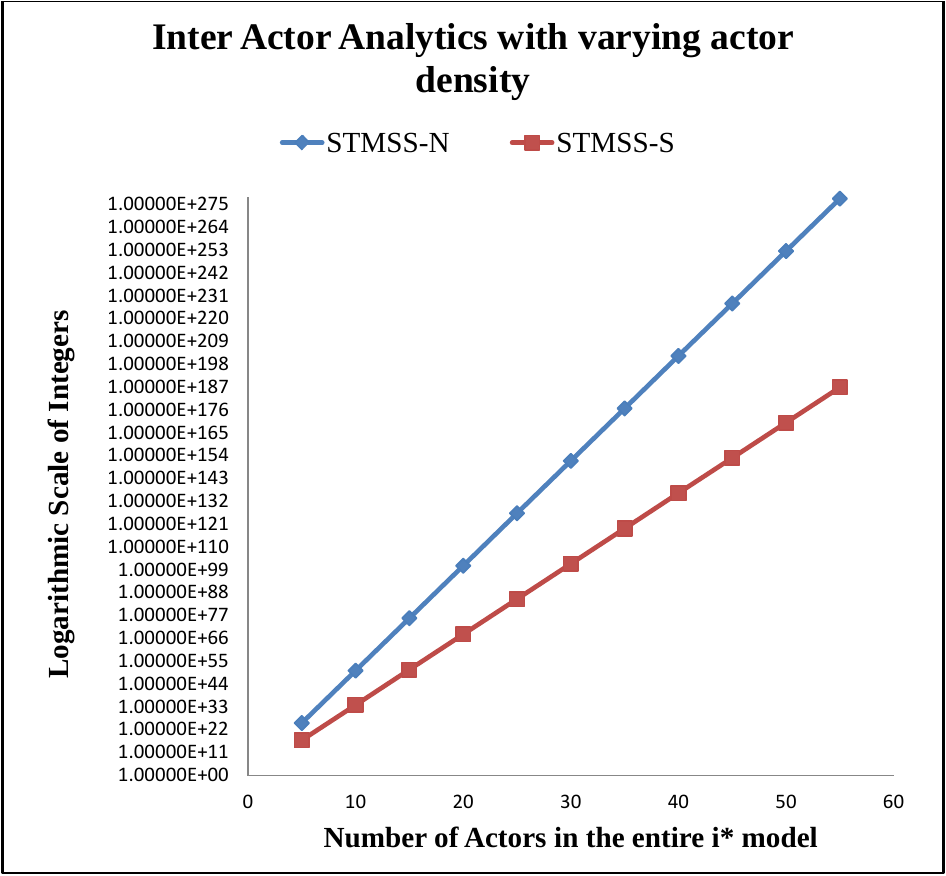}
	\caption{Behaviour analysis with respect to the state transition model space of the entire enterprise for the \textit{Naive Algorithm} (\texttt{STMSS-N}) and the \textit{Semantic Implosion Algorithm} (\texttt{STMSS-S}) as the density of actors in the i$^*$ model varies [\textit{To be reproduced in color on the Web and in black-and-white in print}]}
	\label{fig:IAAnActDen}
\end{figure}

We restrict the number of model elements in each actor to 5 and increase the density of actors from 5 to 55 in steps of 5. Table \ref{table:IAAnActDen} represents such a data set. Figure \ref{fig:IAAnActDen} shows the corresponding graph structure that is obtained by plotting this data.

\begin{table}
	\caption{Inter Actor Analytics obtained by varying Actor Density}
	\centering
		\begin{tabular}{ccc}
		\hline 	                                                       \hline
		\multirow{2}{*}{\begin{tabular}[c]{@{}c@{}}No.\\   of Actors($n$)\end{tabular}} & Naive Algorithm & SI Algorithm \\ \cline{2-3}
		& STMSS$_N$ = $(\frac{(2k)!}{2^k})^n, k=5$         & STMSS$_S$ = $(2520)^n$      \\ \hline
		5     & 1.87528E+25     & 1.01626E+17  \\
		10    & 3.51666E+50     & 1.03277E+34  \\
		15    & 6.59471E+75     & 1.04956E+51  \\
		20    & 1.23669E+101    & 1.06662E+68  \\
		25    & 2.31914E+126    & 1.08396E+85  \\
		30    & 4.34902E+151    & 1.10158E+102 \\
		35    & 8.15562E+176    & 1.11949E+119 \\
		40    & 1.52940E+202    & 1.13768E+136 \\
		45    & 2.86805E+227    & 1.15618E+153 \\
		50    & 5.37840E+252    & 1.17497E+170 \\
		55    & 1.00860E+278    & 1.19407E+187 \\ \hline
	\end{tabular}
	\label{table:IAAnActDen}
\end{table}

Interpretation of the graph is quite intuitive. The blue line represents the growth function of the \textit{Naive Algorithm}. In this case study, it represents the exponential function $(113400)^n$. The red line, on the other hand, plots the growth function of the \textit{Semantic Implosion Algorithm} and represents the exponential $(2520)^n$. With the vertical axis representing a \textit{Logarithmic scale} of integers, the two functions are mapped as straight lines with different gradients. Obviously, the gradient of the blue line is greater than the gradient of the red line. This has the semantic interpretation that the \textit{Naive Algorithm} increases the state transition model space more rapidly as compared to the \textit{Semantic Implosion Algorithm}.

\subsubsection{Variation of the Distribution of Process Elements}
In this particular case study, we fix the number of actors involved in the enterprise i$^*$ model to 5. Keeping the number of actors fixed, the distribution of model elements per actor is increased from 5 to 25 in steps of 5. Assuming uniformity across all the actors in the i$^*$ model, every actor generates it's state transition model space with the same size. The space size changes with varying model element distribution density. Let the size of the state transition model spaces of the individual actors be given by \textit{S$_{1}$, S$_{2}$, ..., S$_{5}$,} respectively, for some model element distribution $k$.

The \textit{Naive Algorithm} combines equations \ref{EQ:Path} and \ref{EQ:EnterpiseSize} to give a function representing the growth rate of the state transition model space as follows:
\begin{equation}
	\label{EQ:IAAnNaive}
\emph{S$_N$} = (\frac{(2k_1)!}{2^{k_1}})^5, \forall k_1, k_1\in\{5,10,15,20,25\}.
\end{equation}

The \textit{Semantic Implosion Algorithm} expands the state transition model space for task decompositions only. Our underlying assumption that there exists a 4-element task decomposition for every group of 5 elements dictates the growth function of the state transition model space as follows:
\begin{equation}
	\label{EQ:IAAnSIA}
	\emph{S$_S$} = (\frac{(2k_2)!}{2^{k_2}})^5, k_2=k_1\div5, \forall k_1, k_1\in\{5,10,15,20,25\}.
\end{equation}

\begin{figure}[h]
	\centering
	\includegraphics[width=0.7\textwidth, page=11]{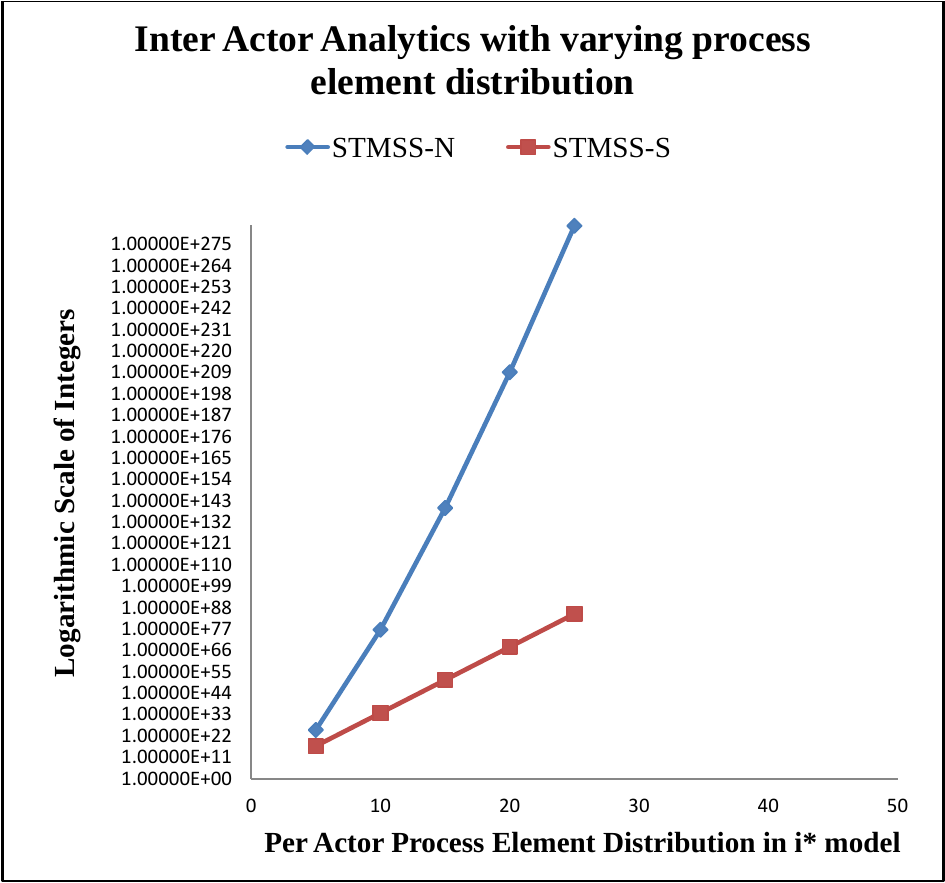}
	\caption{Behaviour analysis with respect to the state transition model space of the entire enterprise for the \textit{Naive Algorithm} (\texttt{STMSS-N}) and the \textit{Semantic Implosion Algorithm} (\texttt{STMSS-S}) as the distribution of model elements within actors in the i$^*$ model varies [\textit{To be reproduced in color on the Web and in black-and-white in print}]}
	\label{fig:IAAnPropElDist}
\end{figure}

The data generated from equations \ref{EQ:IAAnNaive} and \ref{EQ:IAAnSIA} is shown in Table \ref{table:IAAPropElDistr}. The number of actors have been fixed to be 5. Figure \ref{fig:IAAnPropElDist} represents the graph corresponding to this data.

\begin{table}
	\centering
	\caption{Inter Actor Analytics obtained by varying the Distribution of Process Elements}
	\begin{tabular}{ccc}
		\hline                                                         \hline
		\multirow{2}{*}{\begin{tabular}[c]{@{}c@{}}No. of Process \\ Elements ($k_1$)\end{tabular}} & Naive Algorithm & SI Algorithm \\ \cline{2-3}
		& STMSS-N=$(\frac{(2k_1)!}{2^{k_1}})^5$         & STMSS-S=$(\frac{(2k_2)!}{2^{k_2}})^5, k_2=k_1\div5$      \\ \hline
		5     & 1.87528E+25     & 1.01626E+17  \\
		10    & 7.57046E+76     & 1.03277E+34  \\
		15    & 3.47576E+139    & 1.04956E+51  \\
		20    & 2.85249E+209    & 1.06663E+68  \\
		25    & 6.11823E+284    & 1.08399E+85
	\\ \hline
	\end{tabular}
	\label{table:IAAPropElDistr}
\end{table}

The interpretation of the graph is quite similar to the previous graphs. The vertical axis represents a \textit{Logarithmic scale} of integers. Both the exponential function given by equations \ref{EQ:IAAnNaive} and \ref{EQ:IAAnSIA} appear as straight lines. However, the gradients of the two lines are widely different. This implies that the rate of growth of \texttt{STMSS-N} (represented by the blue line) is much greater than that of \texttt{STMSS-S} (represented by the red line).

\subsection{SIA Analytics}
The analytics provided in tables \ref{table:AIAnalytics}, \ref{table:IAAnActDen}, and \ref{table:IAAPropElDistr}, and the corresponding graphs shown in figures \ref{fig:AIAnalytics}, \ref{fig:IAAnActDen}, and \ref{fig:IAAnPropElDist}, all point in the same direction. The obvious conclusion from these data sets is that the \textit{Semantic Implosion Algorithm} provides a huge improvement over the more simple \textit{Naive Algorithm}. The improvement is in the context of space complexity and the \textit{SI Algorithm} provides this improvement with a factor of $10^{15} - 10^{16}$.

Accepting the above conclusion triggers an urge to take an insight into the behaviour of the \textit{SI Algorithm} when both the parameters - \textit{Actor Density} and \textit{Property Element Distribution} - are varied simultaneously. Table \ref{table:SIAAnalytics} provides such a data set. The data are obtained by varying the distribution of model elements in individual actors from 5 per actor to 25 per actor, in steps of 5. The state transition model space size is obtained using the following equation:
\begin{equation}
\emph{STMSS-A}=(\frac{(2k_2)!}{2^{k_2}})^A, k_2=k_1\div5
\label{EQ:SIAAnalytics}
\end{equation}

\begin{figure}[h]
	\centering
	\includegraphics[width=0.7\textwidth, page=13]{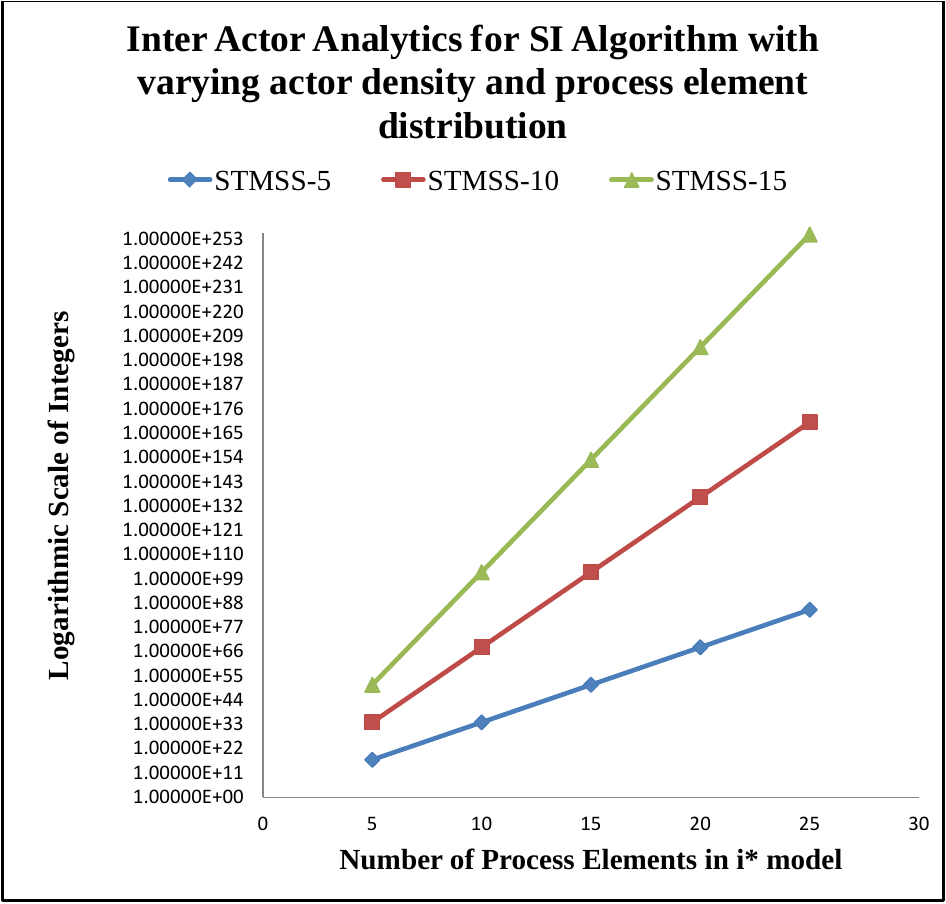}
	\caption{Behaviour analysis of the \textit{Semantic Implosion Algorithm} (w.r.t. the state transition model space) as the distribution of model elements within actors and the actor density in the i$^*$ model are both varied [\textit{To be reproduced in color on the Web and in black-and-white in print}]}
	\label{fig:SIAAnalytics}
\end{figure}

The $A$ in equation \ref{EQ:SIAAnalytics} represents the number of actors. $k_2$ is obtained from $k_1$ as mentioned in the equation due to the assumption that we have a 4-element task decomposition for every group of 5 model elements. Maintaining the uniformity of model element distribution across all the actors of an i$^*$ model, we obtain the data set for 5, 10, and 15 actors, given by STMSS-5, STMSS-10, and STMSS-15, respectively. The graph obtained from the data set in table \ref{table:SIAAnalytics} is shown in figure \ref{fig:SIAAnalytics}.

\begin{table}
	\centering
	\caption{Inter Actor Analytics obtained by varying both Actor Density and Distribution of Process Elements for the \textit{SI Algorithm}}
	\begin{tabular}{cccc}
		\hline \hline
		\multirow{2}{*}{\begin{tabular}[c]{@{}c@{}}No.of Process \\ Elements ($k_1$)\end{tabular}} & \multicolumn{3}{c}{SI Algorithm}          \\ \cline{2-4}
		& STMSS-5     & STMSS-10     & STMSS-15     \\ \hline
			
		5 & 1.01626E+17 & 1.03277E+34  & 1.04956E+81  \\
		10 & 1.03277E+34 & 1.06662E+68  & 1.10158E+102 \\
		15 & 1.04956E+51 & 1.10157E+102 & 1.15617E+153 \\
		20 & 1.06663E+68 & 1.13769E+136 & 1.21349E+204 \\
		25 & 1.08399E+85 & 1.17503E+170 & 1.38069E+255 \\ \hline
	\end{tabular}
	\label{table:SIAAnalytics}
\end{table}

The graph is fairly simple to analyze and interpret. The vertical axis is again a \textit{Logarithmic Scale}. Each of the individual lines (blue, red, and green) are linear, representing exponential growth functions. The fact that the state transition model space size will increase with greater number of actors has already been observed in figure \ref{fig:IAAnActDen}. Hence, the higher positioning of the lines as the number of actors increases. It can also be concluded from figure \ref{fig:IAAnPropElDist} that for a fixed actor density, the state transition model space size increases with increasing density of model elements. Hence, the positive gradient in each of the three lines.

The more important observation here is that the gradient of the lines increases with increasing actor density, i.e., the green line is more steep compared to the red line which, in turn, is steeper than the blue line. We already know that the gradient of the straight lines represents the rate of growth of the exponential functions representing the growth of the respective state transition model spaces. This means that as the actor density increases, the state transition model space increases even more rapidly.

\section{Conclusion}
\label{Concl}
Enterprise architects are aware of the need for temporal information to be captured by a modelling language. However, a requirement specification modelling paradigm like i* is essentially sequence agnostic and rightfully so. i$^*$ models are used to provide an abstract graphical overview of the enterprise to the customer so that he/she has a better understanding of the implications of the requirements as specified by him/her. The true essence of modelling does not reside in providing a graphical interface to the outside world; rather modelling can be exploited for ensuring the correctness of the enterprise being designed by checking the model against inconsistencies, incorrect assertions, counter-possibilities and different other types of anomalies. Formal Model Checking methods and tools exist to achieve this. A correct model can then be used to automate the generation of code snippets that help in the design and testing phases of the System Development Life Cycle.

Enterprise designers are of the same opinion that both Model Checking and Automated Code Generation demand the existence of sequential information within the model. Model Checking tools, typically check a model against certain temporal properties. The need to bridge the gap between i* models and any other business process model is evident. Although model transformations have existed in the industry for quite some time, no work has been done to derive sequential models from i* models. This paper first illustrates and presents a \textit{Naive Algorithm} for extracting sequences from i* model constructs. Simulation results demonstrate how this causes a \textit{hyperexponential explosion} in the state transition model space. The \textit{Semantic Implosion Algorithm} provides an approach to counter this explosion. 

Detailed simulations have been done by applying both the algorithms to similar types of i* models and the results show that the \textit{Semantic Implosion Algorithm} provides a significant improvement over the \textit{Naive Algorithm}. Typically, the state transition model space grows in the order of 10$^{20}$ for the \textit{Naive Algorithm}, whereas, for the \textit{Semantic Implosion Algorithm}, the growth rate is restricted to the order of 10$^3$. Although this may not be the best approach to extract a minimal set of plausible state transition models that can be derived from a given i$^*$ model, it definitely provides a significant improvement over the \textit{Naive Algorithm}. 

The set of possible state transition models, that correspond to a given i$^*$ model, can be further pruned by feeding them into a Model Checking tool like NuSMV and checking them against certain customer-specific temporal properties or compliance rules. All models that generate counter-examples may be discarded. This is one of the biggest advantages of modelling an enterprise. Also, once the set of valid state transition models have been obtained, we can map them to BPMN models, Petri-Nets, or even UML models. This helps Enterprise Architects by allowing the automated generation of code snippets, thereby, reducing the efforts required to build the enterprise. Thus, once the requirements have been finalized and modelled by the architects, the development of the enterprise becomes fully automated, ensuring greater consistency and correctness and reducing the risks of failure.

\section*{Acknowledgement}
This work is a part of the Ph.D. work of Novarun Deb, who is a Research Fellow in the University of Calcutta under the Tata Consultancy Services (TCS) Research Scholar Program (RSP). We acknowledge the contribution of TCS Innovation Labs in funding this research. Part of this work was done by Novarun Deb at the Decision Systems Lab, University of Wollongong, during June-July 2014. We acknowledge the Technical Education Quality Improvement Programme (TEQIP), University of Calcutta, for organizing and sponsoring my visit to the university in Wollongong, Australia.

\bibliography{mybibfile}
\end{document}